\documentclass[12pt,letter]{article}

\usepackage{graphicx}
\usepackage{amsmath}
\usepackage{amssymb}
\usepackage{cancel}
\usepackage[left=1in,right=1in,top=1in,bottom=1in]{geometry}
\usepackage{setspace}
\usepackage{float}
\usepackage[super,sort&compress,comma]{natbib}
\usepackage[font={small}]{caption} 
\usepackage[labelfont=bf]{caption}
\usepackage{bm}
\usepackage[T1]{fontenc}
\usepackage{microtype}
\usepackage{subfigure}
\usepackage{chemformula}
\usepackage[version=4]{mhchem}
\usepackage{textcomp}
\usepackage{color}
\usepackage{mathtools}
\usepackage{bm}
\usepackage{gensymb}
\usepackage{esvect}
\usepackage{booktabs}
\usepackage{multirow}
\usepackage{indentfirst}
\usepackage{mathrsfs}
\usepackage{float}
\usepackage{hyperref}
\usepackage{xcolor}
\usepackage{booktabs}
\usepackage{amssymb} 
\usepackage{bm} 
\newcommand*{\addFileDependency}[1]{
  \typeout{(#1)}
  \@addtofilelist{#1}
  \IfFileExists{#1}{}{\typeout{No file #1.}}
}
\makeatother

\begin{document}


\newcommand{\mub}{\boldsymbol{\mu}}



\title{\textbf{Beyond potential energy surface benchmarking: a complete application of machine learning to chemical reactivity}}
\author{Xingyi Guan$^{1,3}$, Joseph Heindel$^{1,3}$, Taehee Ko$^{4}$, Chao Yang$^{5}$, Teresa Head-Gordon$^{1-3}$\\}
\date{}
\maketitle
\noindent
\begin{center}
$^1$Kenneth S. Pitzer Theory Center and Department of Chemistry, \\
$^2$Departments of Bioengineering and Chemical and Biomolecular Engineering\\
University of California, Berkeley, CA, 94720 USA\\
$^3$Chemical Sciences Division, Lawrence Berkeley National Laboratory, Berkeley, CA, 94720 USA\\
$^4$Department of Mathematics, Penn State University, University Park, PA 16802\\
$^5$Applied Mathematics and Computational Research Division, Lawrence Berkeley National Laboratory\\

corresponding author: thg@berkeley.edu
\end{center}

\begin{abstract}
\noindent
We train an equivariant machine learning model to predict energies and forces for a real-world study of hydrogen combustion under conditions of finite temperature and pressure. This challenging case for reactive chemistry illustrates that ML-learned potential energy surfaces (PESs) are always incomplete as they are overly reliant on chemical intuition of what data is important for training, i.e. stable or metastable energy states. Instead we show here that a “negative design” data acquisition strategy is necessary to create a more complete ML model of the PES, since it must also learn avoidance of unforeseen high energy intermediates or even unphysical energy configurations. Because this type of data is unintuitive to create, we introduce an active learning workflow based on metadynamics that samples a lower dimensional manifold within collective variables that efficiently creates highly variable energy configurations for further ML training. This strategy more rapidly completes the ML PES such that deviations among query by committee ML models helps to now signal occasional calls to the external ab initio data source to further molecular dynamics in time without need for retraining the ML model. With the hybrid ML-physics model we predict the change in transition state and/or reaction mechanism at finite temperature and pressure for hydrogen combustion, thereby delivering on the promise of real application work using ML trained models of an ab initio PES with two orders of magnitude reduction in cost.

\end{abstract}

\section{Introduction}
\label{sec:intro}

\noindent
Machine learning (ML) methods are emerging as an alternative to ab initio molecular dynamics (AIMD) and physical-based potential energy functions (or force fields), once trained on high quality ab initio energies and forces associated with a given conformation of nuclei. Starting with the generalized neural network representation of high dimensional potential energy surfaces (PESs) proposed by Behler and Parrinello\cite{Behler2007}, their work inspired additional state-of-the-art approaches for non-reactive systems, such as ANI-1 for predicting energies of organic molecules\cite{Smith2017,Smith2018,Smith2019}, kernel methods such as Gradient Domain Machine Learning (GDML) that predict atomic forces directly\cite{Chmiela2018,Chmiela2017}, and the introduction of other invariant models based on message-passing architectures developed as generic molecular property predictors\cite{Schutt2018} 
A key development more recently are ML models that are equivariant to translations and rotations through their architectures, showing testing superiority in accuracy benchmarks with greatly reduced quantities of reference data for training.\cite{Thomas2018,Drautz2019,Anderson2019,Qiao2020,Batzner2021,Glick2021,Schutt2021,Unke2021,newtonnet_2022} In what follows we use NewtonNet \cite{newtonnet_2022}, a physics inspired message passing equivariant neural network, as the underlying deep learning model for energy and force prediction for the surprisingly difficult case of hydrogen combustion, for which we developed the HCombustion dataset \cite{HCombustion_data_2022} of energies and forces generated using the $\omega$B97X-V \cite{wb97x-v_2014} DFT functional with the cc-pVTZ basis set. In the case of hydrogen combustion there are at least 19 reaction channels, multiple stable and unstable intermediates that are dependent on a given reactive channel, and complications can arise due to creation of radical species and alternative spin states during the combustion process. 

As such, and regardless of the architecture, the reliability of the ML model still heavily relies on the diversity of the training data, especially for chemically reactive systems that must visit high energy states when undergoing chemical transformations. ML models by their nature interpolate between known training data, but its extrapolation capability is limited, i.e. predictions can be unreliable when molecules or their configurations are dissimilar to those in the training set. Thus, in order to achieve meaningful chemically reactive simulations with ML potentials, the training sets should cover a wide range of structural space with variable energy stability.\cite{Kulichenko_Barros_2023} However, it is challenging to formulate \textit{a priori} a dataset that is balanced and diverse for a given reactive system, and it is not unusual for the ML model to still suffer from an overfitting problem that can lead to models with good accuracy on its original test set but are error-prone when applied to MD simulations, especially in the case for gas phase chemical reactivity in which energy configurations are highly diverse. 

Traditionally, active learning (AL) is a powerful strategy for reducing the amount of labeled data required to develop an ML potential, and selecting informative configurations for labeling.\cite{shapeev_gubaev_1970} Typically AL uses a query by committee strategy, in which variance among a set of identical architectures but stochastically initiated ML models select the most informative data points for labeling, reducing the data generation effort and improving the accuracy of the ML model. However AL informed ML is still not a panacea without having more information. In this study, we propose an active learning workflow for chemical reactivity that utilizes a different information source - namely the sampling efficiencies that are inherent in statistical mechanics methods for rare events. In particular, we have formulated an AL workflow that expands on the originally formulated HCombustion dataset \cite{HCombustion_data_2022} by formulating collective variables (CVs) to first systematically sample a lower manifold of all the intrinsic reaction coordinates of the 19 reactive channels, and then to stochastically sample with metadynamics to  take advantage of its known better ergodicity.\cite{Laio2002,Barducci2008} It is mostly irrelevant whether the CVs are "good", in the sense that the first purpose of the CVs is to trap both the intrinsic reaction coordinate and to sample in the orthogonal directions for much higher energy configurations. Within the smaller subspace metadynamics fills the free energy wells to more rapidly find high energy states such that the ML model learns about configurations whose energies are simply inaccessible, or to find new reactive intermediates which become accessible with temperature and pressure but are of the wrong energy ordering. 

This active learning strategy allows us to reach a final hydrogen combustion ML model that is more diverse and balanced, and more importantly, an energy surface that is relatively smooth. We then utilize the variance of the ML committee models to report back on additional but much fewer PES rough spots, at which we substitute a direct call to an ab initio force to complete a local in time molecular dynamics update, until the ML models recover accurate forces to continue the trajectory, all without further retraining. We illustrate the completeness of the ML hydrogen combustion model with a metadynamics application using well formulated CVs inspired by diffusion maps\cite{Ko2023} to discover the entropic contribution of free energy transition states for hydrogen combustion reaction channels. 

\section{Results}
\label{sec:results}
\subsection{Active Learning workflow}
\noindent
NewtonNet\cite{newtonnet_2022} has been previously trained on the HCombustion dataset\cite{HCombustion_data_2022}(which we refer to as the original dataset) comprised of ab initio molecular dynamics (AIMD) sampling around the intrinsic reaction coordinate (IRC) at 0K near the transition state, as well as systematic normal modes with respect to the IRC to produce PES curvature data, all specific to 19 elemetary reactions (Supplementary Table S1).\cite{Bertels_2020,Li2004-repeat} This yielded a total of $\sim$290,000 potential energies and $\sim$1,260,000 nuclear force vectors. Using $\sim$5k data per reaction, the trained NewtonNet model reaches very good accuracy for energies (mean absolute error (MAE) of 0.14 kcal/mol) and forces (MAE of 0.33 kcal/mol/\AA). 

\begin{figure}[H]
\center
\includegraphics[width=\textwidth]{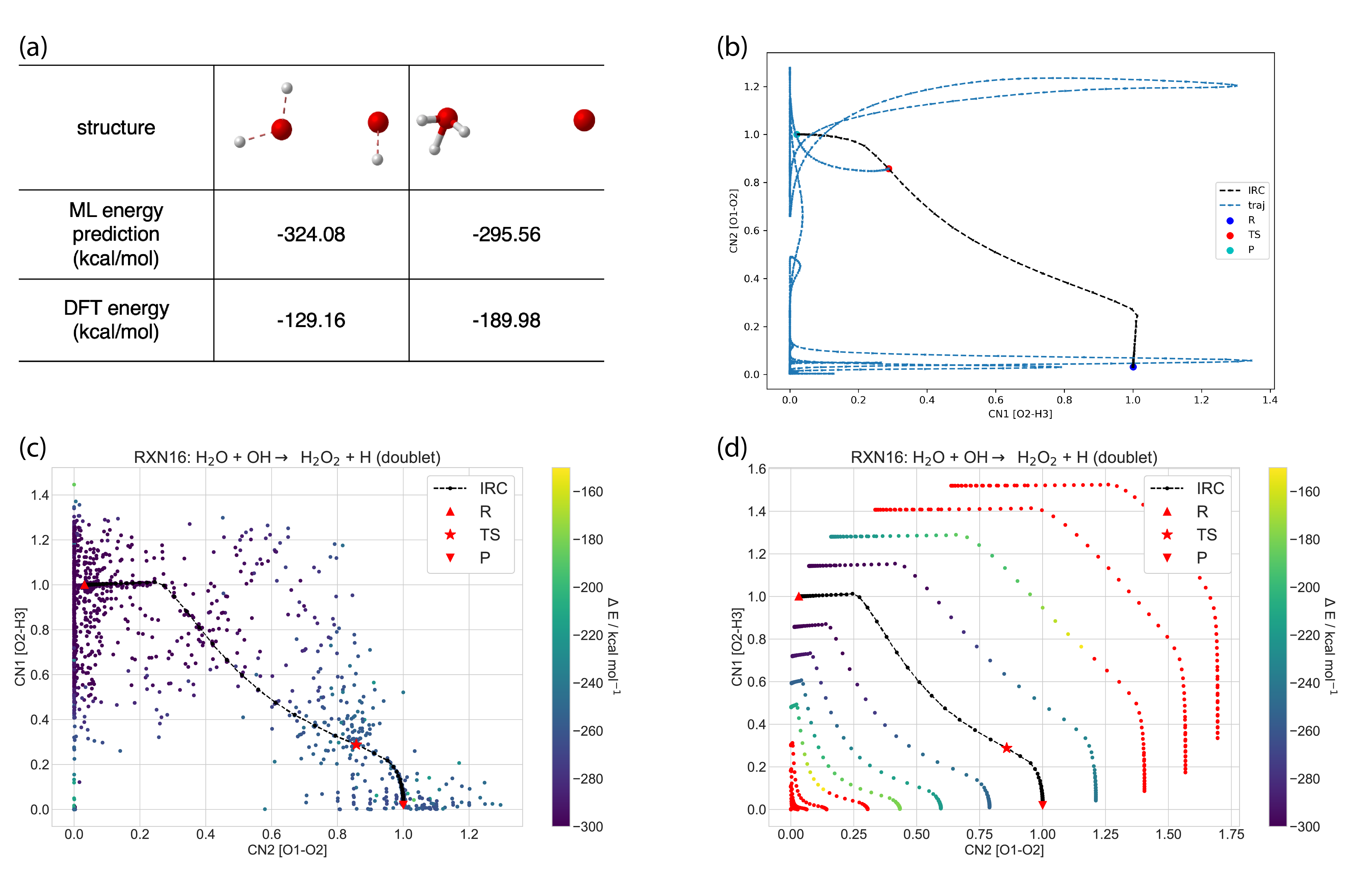}
\caption{(a)Two representative structures that the original ML model predicts with large error. (b) A trajectory for Rxn16 driven with the original ML model visualized on two reaction coordinates CN1 (H4-H5) and CN2(O2-H5). (c,d) Rxn16 DFT potential energy surfaces visualized on two reaction coordinates CN1 (H4-H5) and CN2(O2-H5). (c) the original HCombustion dataset with AIMD and normal mode data (d) the dialation dataset. Each CN represents the formation or breaking of relevant bonds in the reaction process. Energy of each point is color coded in (c) and (d), with red color meaning points with energy higher than the Boltzmann weighting threshold. }
\label{fig:dialation}
\end{figure}

However, when the ML model is applied to molecular simulation, we find it to be highly error-prone, requiring for example new data corresponding to atomization energies to understand the inherent stability of molecules. Furthermore, while the model is trained on relatively stable geometries as well as metastable transition states, it lacks the knowledge of any highly unstable state, and thus predicts configurations with the wrong energy ordering as well as unphysical geometries that were predicted by the L model to be energetically stable. In Figure \ref{fig:dialation}(a), we show representative structures for these two cases; on the left is an unphysical geometry that is close to dissociation, with DFT energy of -129.16 kcal/mol, but the ML model predicts its energy to be as stable as -324.05 kcal/mol. On the right is the hydronium ion that appears in one of the trajectory with an artificially low energy of -295.56 kcal/mol predicted by model, while its actual DFT energy is -189.98 kcal/mol. Figure \ref{fig:dialation}(b) shows such a trajectory, starting from the transition state of rxn16, proceeding through a series of unphysical states, and eventually ending in configurations where all the bonds are broken. 

Hence an MD simulation driven with the model will quickly generate trajectories that go to unphysical regions because the quantum mechanical data will always be small-scale and insufficient for generating a complete potential energy surface. It also emphasizes how unintuitive data acquisition is in regards creating a robust ML model of the PES, especially for chemical reactivity, which requires negative design principles, i.e. to add unphysically high energy species to the training of the ML model. To make this systematic and tractable we formulate an IRC dilation dataset within a lower manifold of collective coordinates that encapsulates the IRC for each reaction channel. Using the example of Rxn16 whose AIMD and normal mode data shown in Figure \ref{fig:dialation}c, each geometry along the IRC curve is proportionally scaled with multiple ratios of CVs to generate new high energy species as shown in Figure \ref{fig:dialation}d. This addition to the dataset helps the models to learn bond dissociation and contraction, as well as spanning a more diverse chemical space than the original dataset.

\begin{figure}[H]
\center
\includegraphics[width=0.5\textwidth]{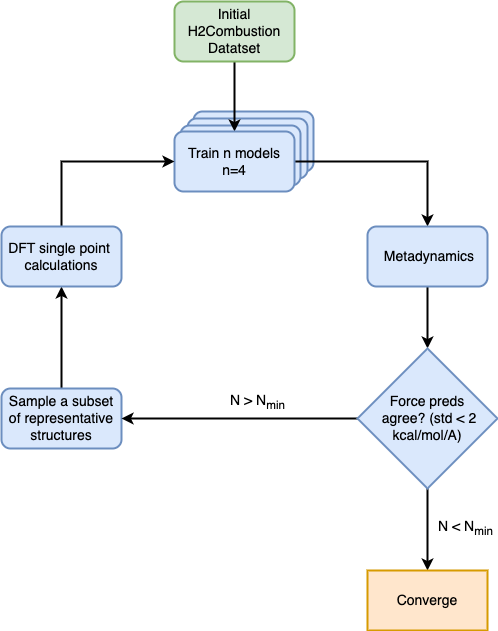}
\caption{Schematic illustration of active learning workflow using query by committee. The four NewtonNet models serve as a starting point of an iterative process, where each round of active learning consists of the following steps: (1) Perform several short metadynamic simulations to explore the configuration space in a lower dimension. (2) When the four models disagree outside standard deviation, collect a representative subset of structures to be included in the training set through downsampling. (3) Perform DFT calculation of energies and atomic forces. (4) Retrain the ensemble of ML models with the updated training set. The details of each of the steps are described in the method section.}
\label{fig:workflow}
\end{figure}

Although the dilation data is helpful, the PES for hydrogen combustion remains incomplete. We thus introduce an active learning workflow in which we trained four models with 1000 structures per reaction selected from the HCombustion dataset and 200 structures per reaction from the dilation dataset using the same architecture but different initial parameters. These four models serve as a starting point of an iterative process, outlined in Figure \ref{fig:workflow}, to systematically improve the ML model through active learning. To allow for a fast turn around time, we use relatively small epoch size and large learning rate in the training through the active learning rounds. More relevant here is that we chose metadynamics, a rare events sampling method, to more efficiently sample previously unseen and unstable structures through an external biasing potential that forces the system to explore regions of high (free) energy. While in the usual context poor selection of the low-dimensional descriptors affects the rate at which transitions are enhanced, in this particular context we are using metadynamics as a tool to fill in the holes in the ML PES in which the goodness of the CVs is less important than the ability to sample diverse conformations of high energy variance.  


\begin{figure}[H]
\center
\includegraphics[width=0.9\textwidth]{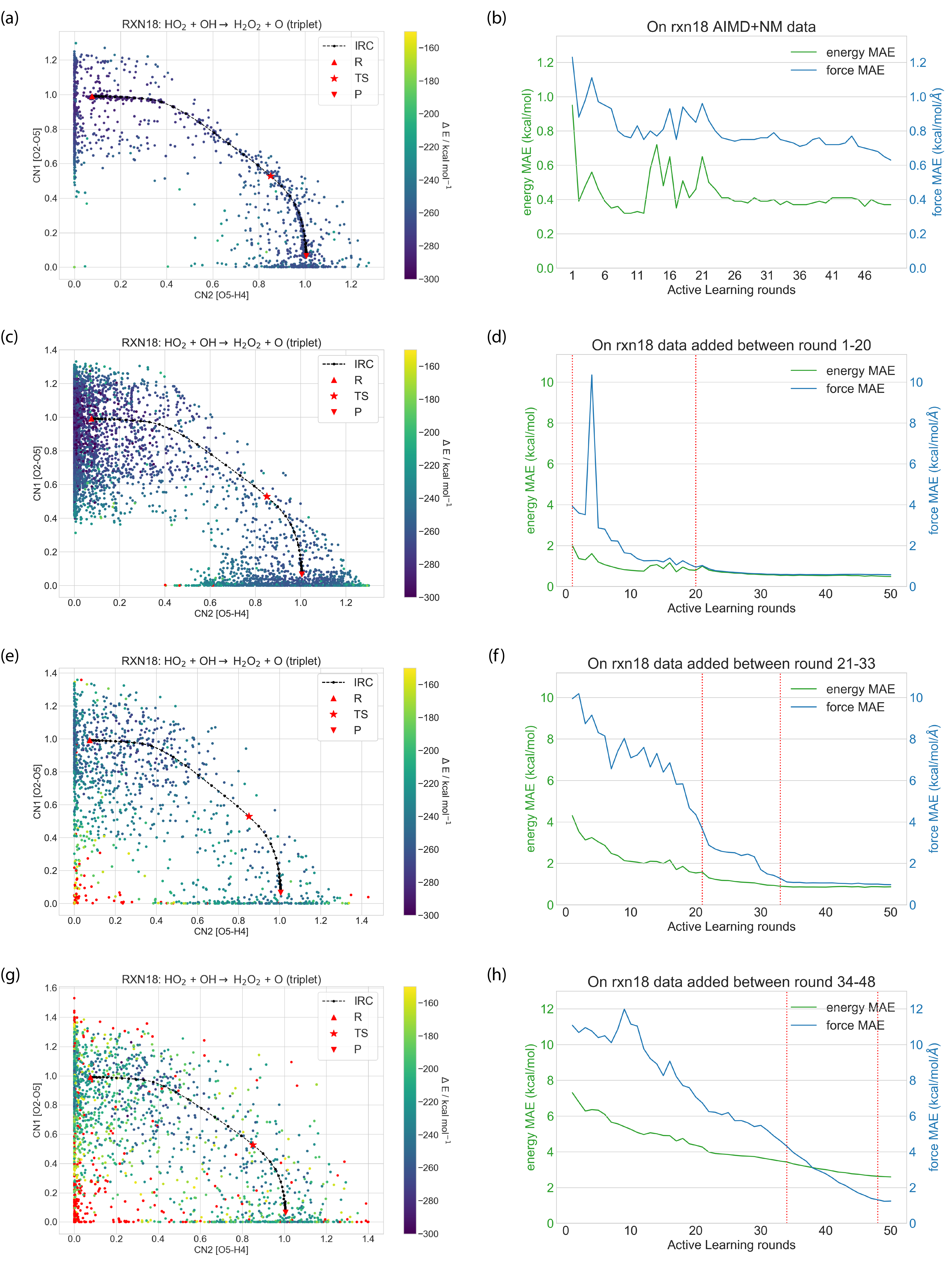}
\caption{Potential energy surface in collective coordinates (left) and change in energy and force mean absolute error(MAE) as active learning round proceeds (right) for Reaction 18. The potential energy surface projected onto two reaction coordinate CN(O2-O5) and CN(O5-H4). Energy of each point is color coded, with red color meaning points with energy higher than the Boltzmann weighting threshold. (a,b) Test set of rxn18 in the original dataset composed of data generated from AIMD and normal modes calculations. (c,d) Data added between active learning round 1-20, sampled with 2000fs metadynamics. (e,f) Data added between active learning round 21-33, sampled with 5000fs metadynamics. (g,h) Data added between active learning round 34-48, sampled with 10000fs metadynamics.}
\label{fig:al_result_18}
\end{figure}

To illustrate the active learning approach, we consider Rxn18 as an example in which the potential energy surface is projected onto two reaction coordinates CN(O2-O5) and CN(O5-H4), with results shown in Figure \ref{fig:al_result_18}. The ML model performance was tracked by analyzing both the original data points derived from AIMD and normal modes calculations,

\noindent
and the newly added data points accumulated during the active learning procedure. Because we used longer metadynamics simulations for sampling as the active learning rounds proceeded and as errors decreased, we show this by dividing the active learning data into four batches: the original data derived from normal mode and AIMD near the transition state of the IRC (Figure \ref{fig:al_result_18}(a,b)), data sampled with 2000fs metadynamics between active learning round 1-20 (Figure \ref{fig:al_result_18}(c,d)), data sampled with 5000fs metadynamics between active learning round 21-33 (Figure \ref{fig:al_result_18}(e,f)), data sampled with 10000fs metadynamics between active learning round 34-48 (Figure \ref{fig:al_result_18}(g,h)). 

As shown on the left side of Figure \ref{fig:al_result_18}, active learning spans more of the reaction coordinate space and in favor of learning high energy states, which indicates that our sampling method is effective in learning points to avoid in the MD trajectory. The lack of points in the upper right quadrant in early stages of active learning indicates that the dilation dataset was sufficient (Figure \ref{fig:al_result_18}c), and that the model became more accurate near the reaction pathway basins since there is little sampling in this region during later rounds of the active learning process using metadynamics (Figure \ref{fig:al_result_18}(e,g)). The right side of Figure \ref{fig:al_result_18} shows that there is a very significant improvement in model error when predicting on the newly added data. Using the original ML model these new sampled points from active learning created large energy and force errors as high as $\sim$8 kcal/mol and $\sim$12 kcal/mol/\AA \, respectively, which of course would create an untenable MD trajectory if those geometries are visited during an actual application. But after $\sim$50 rounds of active learning, the model prediction on new geometries are hugely improved, with 0.97 kcal/mol/\AA \ error in forces on the data points collected between round 21-33, and 1.24 kcal/mol/\AA \ force error on the data points collected between round 34-48. 

The somewhat larger errors are still excellent given the much larger energy range required to generate physically meaningful trajectories, which derives from the more significant improvements of the new data points added with the active learning workflow. It also reflects another important strategy in active learning, which is that high energy configurations must be learned but they do not have to be as accurate, and this can be reflected in the loss function through Boltzmann weighting (see Methods). This is supported by the fact that the original data points are still predicted with relatively small errors in energy and forces with the active learning model (~0.4 kcal/mol energy MAE and ~0.8 kcal/mol/\AA \ force MAE).  

\subsection{Committer Analysis and the Free Energy Surface}
\noindent
To investigate the outcome of the active learning to create an ML potential for hydrogen combustion, we performed committer analysis on the elementary reaction channels for hydrogen combustion, starting the trajectories from the IRC transition state of the reaction with temperature set to 500 K. Diatomic reactions (rxn05, 06, 07 and 08) and barrierless reaction 15 were not considered. Reaction 12 has two spin states doublet and quartet in the original dataset, and only the transition state with lower energy (12b quartet) was considered because the model was trained with energy and forces from the lower energy spin state. 

Table \ref{tab:committer} presents the committer results for both the original model without active learning and the final model after all active learning rounds. The final ML model gives quite different committer statistics than the original model, in which the AL model shows a more even chance to commit to reactant and product for a majority of the reactions. Hence although the AIMD and NM sampling provided data representation in the tube around the IRC, it is insufficient such that the commitor analysis is qualitatively different in a majority of the reactions. But once the model is more complete using our AL strategy, there is a shift in the forward/backward committer distribution in most of the reaction channels in which the 0 K IRC is still a good estimator. Even so, at 500 K there is some evidence of a non-negligible entropy component to the free energy transition state. 

\begin{table}[h]
\centering
\caption{Committer statistics using (a) the original model (b) the final model after active learning rounds}
\label{tab:committer}
\begin{tabular}{|c | c c | c c|}
\hline
  & \multicolumn{2}{|c|}{\textbf{Original Model}}  & \multicolumn{2}{|c|}{\textbf{Final Model}}\\
\hline
\textbf{rxn} & \textbf{reactant(\%)} & \textbf{product(\%)} & \textbf{reactant(\%)} & \textbf{product(\%)} \\ 
\hline
01 & 12 & 88 & 25 & 75\\
02 & 0 & 100 & 44 & 56\\
03 & 48 & 49 & 50 & 50\\
04 & 100 & 0 & 44 & 56\\
09 & 86 & 14 & 62 & 38\\
10 & 58 & 41 & 55 & 45\\
11 & 91 & 9 & 55 & 45\\
12 & 28 & 72 & 45 & 55\\
13 & 55 & 44 & 65 & 35\\
14 & 7 & 93 & 40 & 60\\
16 & 0 & 100 & 48 & 52\\
17 & 96 & 4 & 49 & 51\\
18 & 22 & 78 & 43 & 57\\
19 & 95 & 5 & 52 & 48\\
\hline
\end{tabular}
\end{table}

With a more complete ML PES obtained through the active learning procedure, we ran long metadynamics trajectories and reconstructed their free energy surface to determine the transition state free energy for reactions 9, 13, 14, and 18. Unlike in the exploration phase where any CV can help discover new structures, in the real application to reconstruct the free energy surface, a good choice of CVs is important. Due to the diffusive nature of gas phase reactions, the distance between two fragments can be arbitrary and lead to convergence problem if we directly use distance as the collective variable. Here we use CVs inspired by our recent study of diffusion maps in order to assess CVs suitable for each reaction as evaluated through their correlations with the diffusion coordinates. We found that the coordination number(CN) between atom i and j:
\begin{equation}
    CN(ij) = \frac{2}{1+exp(3(r_{ij}-r_{eq}))}
\end{equation}
is often a good CV, where $r_{eq}$ is the equilibrium distance between i and j, usually selected as the equilibrium bond length unless otherwise specified. Restraints on certain bond distance and bond angles are also necessary to keep the metadynamics trajectory inside the relevant chemical space defined by the CVs. 

\begin{figure}[H]
\center
\includegraphics[width=0.8\textwidth]{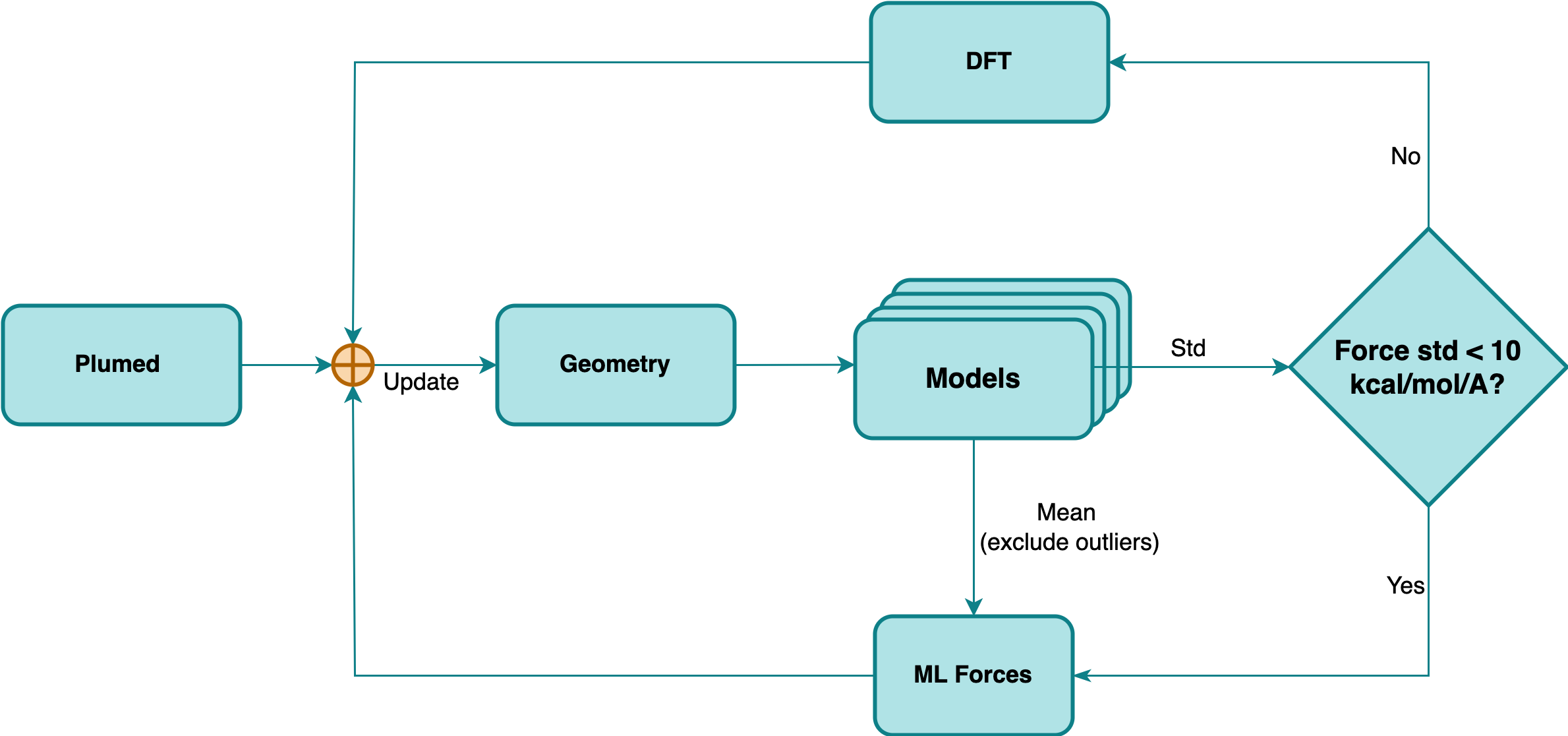}
\caption{Schematic illustration of the new workflow for rebuilding the free energy surface. The metadynamics proceeds with a model that utilizes query by committee ML trajectories to signal when forces have degraded. In lieu of retraining, the AIMD forces are directly substituted to complete the MD time step.}
\label{fig:metad_production}
\end{figure}

We also exploit the strength of having an ensemble of ML models in which the standard deviation between the models measures the reliability in the prediction of ML forces on a given configuration. We therefore devise a hybrid mode of running a simulation in which we use ML forces when the standard deviation among models is relatively small, and with ab initio forces at the same level of theory used for ML training when the model predictions are not as reliable (Figure \ref{fig:metad_production}). 


As we support below, this hybrid mode can be (1) much less expensive than fully retraining the ML models with the new DFT data, and (2) will be more accurate and maintain stability when uncertainties arise, although with appropriate parallelization one could imagine a constant retraining while the MD trajectory proceeds with the hybrid model. Ultimately this involves a tradeoff between how complete is the active learning process vs. The relative cost of the ML and DFT forces; in our case on average it takes $\sim$0.11 seconds for a ML force step and $\sim$12 seconds to generate the \textit{ab initio} forces. 

\begin{figure}[H]
\center
\includegraphics[width=0.925\textwidth]{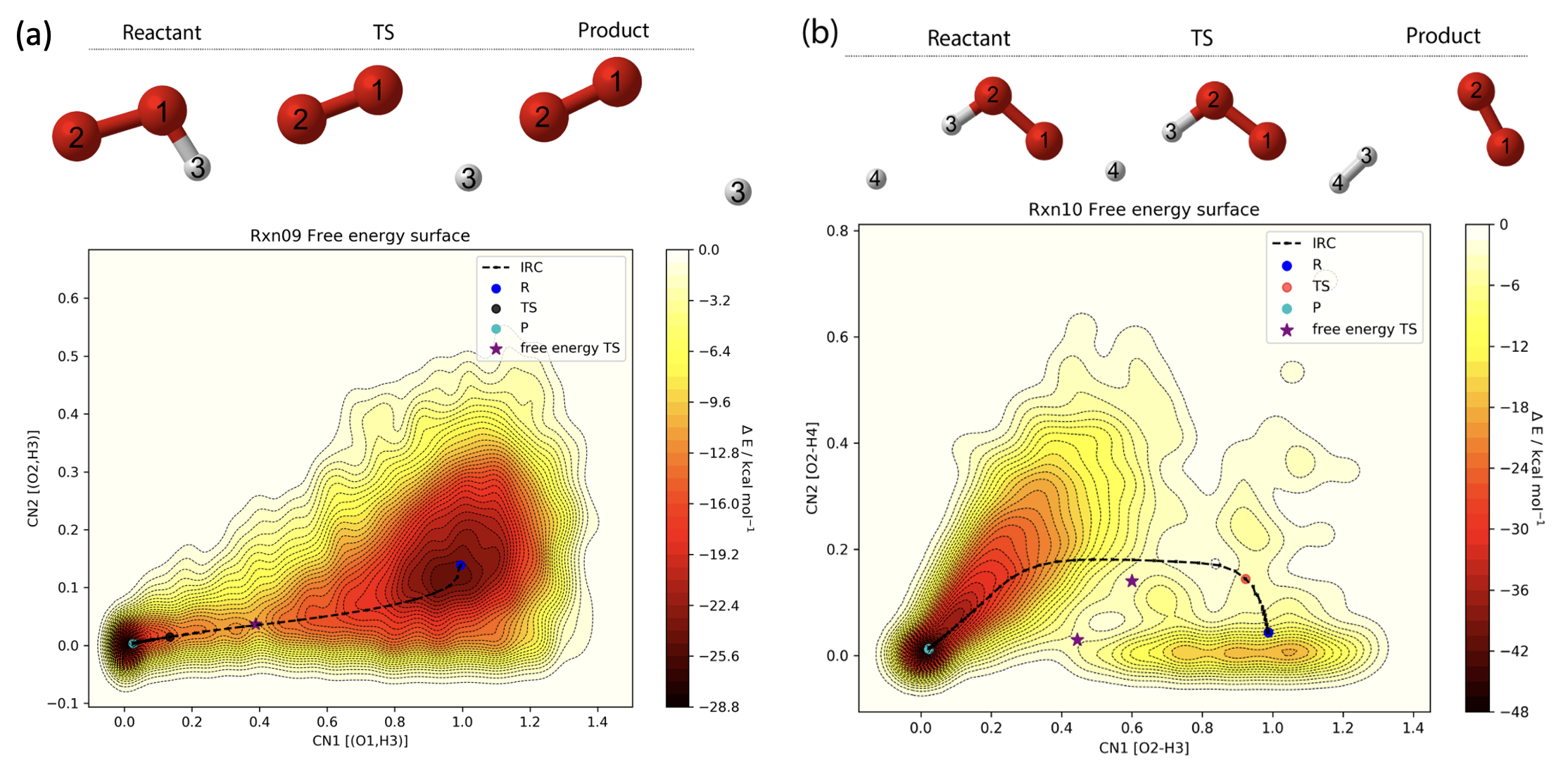}
\includegraphics[width=0.925\textwidth]{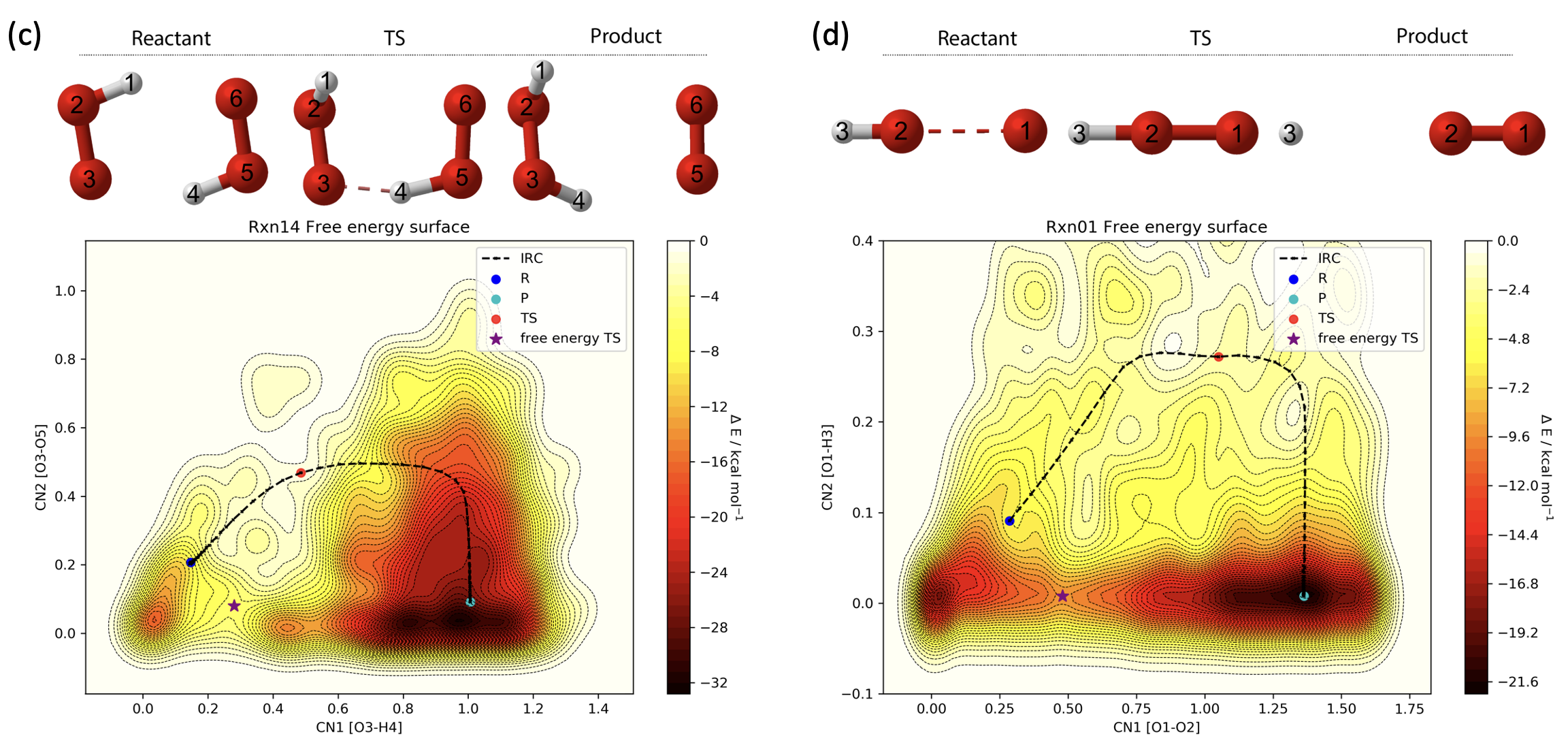}
\caption{Free energy surface reconstructed from metadynamics for hydrogen combustion. (a) Reaction 9, (b) Reaction 10, (c) Reaction 14, and (d) Reaction 1. The original IRC pathway is labeled with a red dot for the transition state and the free energy transition state is labeled as a purple star.}
\label{fig:FES}
\end{figure}

Figure \ref{fig:FES} shows the reconstructed free energy surfaces using the hybrid model for the more interesting reaction channels where the reactant-to-product ratios are uneven. In Rxn 9, when we use CN(O1-H3) and CN(O2-H3) as the collective variables we find that the IRC transition state leans towards the reactant side on the free energy surface, consistent with the 62:38 committer ratio, and thus shifts quite significantly on the free energy surface toward the product well (Figure \ref{fig:FES}a). For Rxn 10 we found CN(O2-O3) and CN(O2-H4) to be good collective variables, where CN(O2-O3) monitors the reaction progress and CN(O2-H4) helps separates the transition state from the stable wells. Consistent with the 55:45 reactant-to-product committer ratio the IRC transition state leans towards the reactant side, but on the free surface two possible transition pathways are evident (Figure \ref{fig:FES}b). One pathway is very similar to the IRC but with the transition state moving closer to the product well, while the other pathway demarcated by a smaller CN(O2-H4) value involves breaking the O2-H3 and forming the H3-H4 molecule in an asynchronous manner.

The IRC transition state of both Rxn14 and Rxn01 lean very heavily toward the product side on the free energy surface (Figure \ref{fig:FES}c,d), as is consistent with the much more skewed commitor reactant-product ratios in Table 1. Accordingly, the free energy transition state shifts dramatically in both reactions, with saddle point regions that exhibit very large free energy stabilization of 8-12 kcal/mole relative to the IRC transition state arising from entropic effects. As a control we tested Rxn 17 and Rxn 18 in which the reactant-to-product ratio in the committer analysis is close to 50-50. In both cases the IRC and free energy transition state are consistent indicating that the coordination numbers tend to be good collective variables and that the entropic factors governing these reactions are small. 

Through this process, we find that the number of unreliable region of PES are relatively infrequent and the deviations are small enough such that energy conservation is not badly effected. For the Rxn 18 trajectory the full DFT force was called 1364 times over a total of 1 million steps, which means 99.86\% of the trajectory is driven by the ML model, and only 0.14\% of steps is calculated through the \textit{ab initio} update. For the Rxn10 trajectory the DFT force was called only 136 times, which means that 99.99\% of the total MD steps were generated by the ML forces.  The small chance to go through ab initio updates reflects that the learned PES after our negative design and metadynamics active learning procedures is largely complete, especially for Rxn 10. But it is important to emphasize that without the \textit{ab initio} calls the MD trajectories would become corrupted beyond repair and this hybrid ensemble approach allows for highly efficient simulation with a time scale similar to the pure ML run but with more stability, offering a preventative and alternative to occasional model hallucinations.

\section{DISCUSSION AND CONCLUSIONS}
\noindent
Machine learning (ML) methods have traditionally aimed to train a faithful surrogate model of ab initio energy and forces for different configurational arrangements of a system in order to explore reaction chemistry more efficiently than \textit{ab initio} molecular dynamics. For real applications it is likely that the interpolative nature of machine learning will always suffer from data insufficiencies that will limit its total replacement of physics based methods. However, this work emphasizes that we can exploit multiple hybrid model strategies to accelerate chemical predictions that will be reliable, especially in regards chemical reactivity. The first recognition is that we need a negative design aspect to data acquisition. Although there has been some focus on collecting data on possible intermediates or in the metastable regions of the potential energy surface, in fact very unphysical states also should be part of the data. But nearly all reference datasets are insufficient as they lack unusual and/or high energy configurations because they are not considered to be chemically relevant. 

To uncover this more unintuitive data, we devise an active learning workflow  where we use enhanced sampling within a lower manifold of collective coordinates along with a query by committee stategy to systematically improve the ML PES in its uncertain region. By conscious negative design through the dialation approach and more diverse sampling with metadynamics, we achieved a more complete and robust ML representation of the hydrogen combustion reactive PES. Metadynamics, as a rare event sampling method, turns out to be an efficient sampling tool for unstable structures, which helps the AL workflow to identify holes in the PES landscape to inform the ML model through retraining with such data. Using metadynamics only as a sampling tool, the tricky CV selection step can be avoided by starting with any reasonable or intuitive CVs. 

Ultimately to address real world application studies using ML potentials, any \textit{a priori} formulated reference dataset, even with active learning approaches, will be too small-scale for true PES completeness. In this case any large variance among ML models provides a nice strategy to invoke a hybrid model, which smooths over rough spots on the ML PES with calls to the ab initio data source. This hybrid ML-AIMD model allowed us to reconstruct and perform analyses on free energy surfaces for many of the reaction channels for hydrogen combustion with much greater accuracies, but at hoped for ML computational efficiencies reaching two orders of magnitude improvement over AIMD.

One alternative approach could be delta learning that learns the difference between a lower level theory and a higher level theory. This type of simulation would require calling both the lower theory method and the ML model at each and every step, and combine them to mimic the PES of the higher level theory. This approach can avoid the detrimental effect from ML rough spots by making ML correction zero whenever the ML models disagree. However, the quality and cost of this type of simulation would highly depend on the underlying lower level theory method, and switching between the two surface can affect the energy conservation in the trajectory. We have explored the idea of using a physical model (C-GeM \cite{Guan_Leven_2021}) to represent the long range behavior and use ML to delta-learn the rest of the interaction, but found that the intrinsic error from the C-GeM model adds challenge for ML models to learn the more complex pattern. At such, we give up on this delta learning idea and decide to stick with the hybrid model we devise.

\section{METHODS}
\label{sec:methods}

\subsection{Dialation Data Preparation}
\noindent
To address the problem that the initial Dataset from Ref. \cite{HCombustion_data_2022} is missing high energy states, we prepared some additional data by proportionally scaling each geometry in the IRC is with multiple ratio (0.6, 0.7, 0.8, 0.9, 1.1, 1.2, 1.3, 1.4 , 1.6, 1.8, 2.0, 2.4, 2.8, 3.2). The energy and forces of these new geometries were obtained with QChem\cite{qchem_2014} force calculation using range separated hybrid meta-GGA functional $\omega$B97X-V \cite{wb97x-v_2014} with the cc-pVTZ basis set.

\subsection{Active Learning: Query by Committee}
\noindent
In the field of molecular modeling, active learning has found another valuable application for reducing ML errors through improving data selectivity and efficiency. Here we exploit the query by committee idea to find a label the most informative data points through this iterative process: 
\begin{enumerate}
\item Perform short metadynamic simulations to explore the configuration space in a lower dimension. 
\item When the four models disagree outside standard deviation, collect a representative subset of structures to be included in the training set through downsampling. 
\item Perform DFT calculation of energies and atomic forces to label the new data. 
\item Retrain the ensemble of ML models with the updated training set. 
\end{enumerate}
The details of each of the steps are described in the following subsections.

\subsection{Training of the NN PES}
\label{subsec:training}
\noindent
The NewtonNet model was used to train the NN potential in this work. Details of this method can be found in Ref. \cite{newtonnet_2022}. The initial four models are trained for 2000 epochs with 1000 data points per reaction from the original dataset with addtional 200 data points from dialation data, initialized with different weights. The following models are trained with exactly the same setting but with additional data sampled from short metadynamics simulations. The cutoff radius was set to 5.0 \AA. The initial learning rate was set to 0.001 and with 0.7 learning rate decay. The loss is defined as

\begin{equation}
    \begin{split}
    \mathcal{L} & = \frac{\lambda_{E}}{M} \sum_{m}^{M} w_m \left(\widetilde{E}_{m}-E_{m}\right)^{2} \\ \nonumber
    &   + \frac{\lambda_{F}}{M} \sum_{m}^{M} \frac{ w_m }{3 N_{m}} \sum_{i}^{N_{m}}\left\|\widetilde{\bm{F}}_{m i}-\bm{F}_{m i}\right\|^{2} \\ \nonumber
    &   + \frac{\lambda_{D}}{M \times N_{m}} \sum_{m}^{M}  w_m \sum_{i}^{N_{m}}\left(1-\frac{{\bm{\mathcal{F}}}_{m i}^{L} \cdot \bm{F}_{m i}}{\left\|{\bm{\mathcal{F}}}_{m i}^{L}\right\|\left\|\bm{F}_{m i}\right\|}\right)
    \end{split}
\end{equation}
where the prefactor of the energy error $\lambda_E$ is set to 1, the prefactor of force error $\lambda_F$ is set to 20, and the prefactor $\lambda_D$ for latent force direction is set to 1. 

We also use the following Boltzmann weighting factor $w_m$, defined as
\begin{equation}
 w_m = 
 \begin{cases}
    1     &\text{ if } E_m <= E_{thresh}
    \\
    exp({\frac{(-(E_m-E_{thresh})}{k_BT})}) & \text{ if } E_m > E_{thresh}\nonumber,
    
 \end{cases}
\end{equation}
to bias the training towards data point within a relevant energy scale. $E_{thresh}$ is a per-atom quantity that put less weighting on all data points with energy higher than this number. It is chosen to be 16.744 kcal/mol/atom for the Hydrogen combustion system, which is 10 kcal/mol/atom higher than the highest per atom energy among all reaction channels in IRCs. To completely converge the ML model, we added one final training step with all previously added data, using a larger epoch size (5000 steps), and a more patient learning rate decay to give the final model that we later use for determining predictions on the free energy surface.

\subsection{Metadynamics}
\noindent
New structures are sampled through short metadynamics trajectories for 6 reaction channels: rxn09, rxn10, rxn13, rxn16, rxn17 and rxn18. For each step in the metadynamics simulation, the atomic forces are evaluated by four NN PES models simultaneously. Outlier among the 4 predictions is removed if the absolute difference between the outlier in question and the closest number is larger than the 95\% confidence limit value of the Dixon Q's test. Then the mean of model predicted forces is modified by plumed \cite{plumed_2014} to allow for enhanced sampling. In this work, all enhanced sampling simulations are performed with the well-tempered metadynamics, in which a Gaussian centered at the visited point is periodically added to the potential.  The simulation is driven through the atomic simulation environment(ASE)\cite{ase_2017} with a specifically tailored calculator that provide energy and forces for a given structure through the above protocol. The simulation is conducted at 300K with increasing length as we have more active learning rounds: 2ps between active learning round 1-20, 5ps round 21-33 and 10ps between round 34-48. 

During the simulation, standard deviation on atomic forces over an ensemble of NN potentials is monitored. Whenever the maximum over the atoms exceeds a predefined threshold (2 kcal/mol/$\text{\AA}$), the configuration is selected for further downsampling. 

\subsection{Clustering and Down Selection}
\noindent
Each molecule is first represented as a Coulomb matrix\cite{rupp_2012} that includes the nuclear charges (Zi and Zj) of atom i and j along with their Cartesian coordinates (Ri and Rj). 
\begin{equation}
    \boldsymbol{C}_{ij} = \begin{cases}
    0.5 Z_i^{2.4}, i = j
    \\
    \frac{Z_iZ_j}{\lvert\boldsymbol{R_i}-\boldsymbol{R_j}\rvert}, i \neq j
    
 \end{cases}
\end{equation}
To reduce the dimension of the dataset while retaining the majority of structural information, the Coulomb matrix was transformed into the eigen-spectrum by solving the eigenvalue problem $\boldsymbol{Cv} = \lambda\boldsymbol{v}$ subject to the constraint $\lambda_i \geq \lambda_{i+1}$. The Mini Batch KMeans clustering algorithm was then applied to categorize the sub-datasets into smaller clusters based on the eigen-spectrum. The value of K (the number of clusters) is chosen automatically with a scaled inertia approach \cite{herman-saffar_2021}. The scaled inertia is formulated as
\begin{equation}
    \text{Scaled Inertia} = \frac{I(K)}{I(K=1)} + \alpha K
\end{equation}
where the inertia(I) is the sum of squared distance of samples to their closest cluster center: 
\begin{equation}
    I(K) = \sum_{i=1}^N{(x_i-C_k)^2}
\end{equation}
where N is the number of samples and $C_k$ is the centroid of a cluster.  $\alpha$ is a manually tuned factor that gives penalty to the number of clusters, here we chose $\alpha =0.0002$. We chose the K value that gives the minimum scaled inertia among all K<300 to do the mini batch K-means clustering on all molecules from a given reaction channel. Afterward, we randomly picked a structure from each cluster, whose energy and forces will be calculated and then be included into the new training set. The final data set had 48,582 data points sampled in the AL procedure.

\subsection{DFT single point force calculations}
\noindent
The structures that passes the clustering and downselection process are gathered for labeling and retraining. They are labeled through DFT force calculations through QChem using $\omega$B97X-V \cite{wb97x-v_2014} functional with the cc-pVTZ basis set. Some of the structures are tricky to deal with due to the diffusive nature of gas phase simulation. Therefore, for all structures with atomic oxygen that is separated from the rest of the structure for more than 2 \AA apart, we introduced and additional run with FRAGMO initial guess for SCF calculation. \cite{khaliullin_2007} We compared the final energy obtained from a calculation with fragmo initial guess and the one with the usual superposition of atomic densities(SAD) inital guess, and take whichever one with lower energy to give the final energy and forces for labeling the data.

\subsection{Reconstructing free energy surface with metadynamics}
\noindent
The free energy surfaces were calculated from longer metadynamic simulations. The Langevin thermostat was used to maintain temperature at 300K, with a friction coefficient of 0.002 a.u. In the metadynamics simulations, the Gaussians adopted have an initial height of 5 kJ/mol and width of 0.05 for the CVs. A Gaussian was deposited every 100 step with a bias factor equal to 10. The simulations were 200 ps long with step size of 0.2 fs.
The free energy were calculated using the sum\_hills utility in Plumed\cite{plumed_2014} and its surface plotted with python matplotlib.

\section{ACKNOWLEDGMENTS}
\noindent
We thank the CPIMS program, Office of Science, Office of
Basic Energy Sciences, Chemical Sciences Division of the U.S. Department of Energy under Contract DE-AC02-05CH11231 for support of the machine learning approach to hydrogen combustion. This work was supported by the U.S. Department of Energy, Office of Science, Office of Advanced Scientific Computing, and Office of Basic Energy Sciences, via the Scientific Discovery through Advanced Computing (SciDAC) program for the collective variables. This work used computational resources provided
by the National Energy Research Scientific Computing Center (NERSC), a U.S. Department of Energy Office of Science User Facility operated under Contract DE-AC02-05CH11231.

\section{DATA AND CODE AVAILABILITY}
\noindent
Coordinates of geometries used in the training: https://doi.org/10.6084/m9.figshare.23290115.v1. The full workflow code can be found in https://github.com/THGLab/
H2Combustion\_AL.

\section{SUPPORTING INFORMATION}
\noindent
We provide a table describing the hydrogen combustion reaction channels 1-19, and the amount of data accumulated over active learning rounds.

\noindent

\section{DECLARATION OF INTERESTS}
\noindent
The authors declare no competing interests.

\section{AUTHOR CONTRIBUTIONS}
\noindent
X.G. and T.H.G. designed the project. X.G. carried out the AIMD simulations, metadynamic calculations, and active learning. X.G. and T.H.G designed the collective coordinates with the help of J.H.,T.K.,C.Y. All authors discussed the results and made comments and edits to the manuscript.

\bibliography{references}

\begin{thebibliography}{10}
\expandafter\ifx\csname url\endcsname\relax
  \def\url#1{\texttt{#1}}\fi
\expandafter\ifx\csname urlprefix\endcsname\relax\def\urlprefix{URL }\fi
\providecommand{\bibinfo}[2]{#2}
\providecommand{\eprint}[2][]{\url{#2}}

\bibitem{Behler2007}
\bibinfo{author}{Behler, J.} \& \bibinfo{author}{Parrinello, M.}
\newblock \bibinfo{title}{Generalized neural-network representation of
  high-dimensional potential-energy surfaces}.
\newblock \emph{\bibinfo{journal}{Physical Review Letters}}
  \textbf{\bibinfo{volume}{98}}, \bibinfo{pages}{146401}
  (\bibinfo{year}{2007}).
\newblock
  \urlprefix\url{https://link.aps.org/doi/10.1103/PhysRevLett.98.146401}.

\bibitem{Smith2017}
\bibinfo{author}{Smith, J.~S.}, \bibinfo{author}{Isayev, O.} \&
  \bibinfo{author}{Roitberg, A.~E.}
\newblock \bibinfo{title}{Ani-1, a data set of 20 million calculated
  off-equilibrium conformations for organic molecules}.
\newblock \emph{\bibinfo{journal}{Scientific Data}}
  \textbf{\bibinfo{volume}{4}}, \bibinfo{pages}{170193} (\bibinfo{year}{2017}).
\newblock \urlprefix\url{https://doi.org/10.1038/sdata.2017.193}.

\bibitem{Smith2018}
\bibinfo{author}{Smith, J.~S.}, \bibinfo{author}{Nebgen, B.},
  \bibinfo{author}{Lubbers, N.}, \bibinfo{author}{Isayev, O.} \&
  \bibinfo{author}{Roitberg, A.~E.}
\newblock \bibinfo{title}{Less is more: Sampling chemical space with active
  learning}.
\newblock \emph{\bibinfo{journal}{The Journal of Chemical Physics}}
  \textbf{\bibinfo{volume}{148}}, \bibinfo{pages}{241733}
  (\bibinfo{year}{2018}).
\newblock \urlprefix\url{https://doi.org/10.1063/1.5023802}.

\bibitem{Smith2019}
\bibinfo{author}{Smith, J.~S.} \emph{et~al.}
\newblock \bibinfo{title}{Approaching coupled cluster accuracy with a
  general-purpose neural network potential through transfer learning}.
\newblock \emph{\bibinfo{journal}{Nature Communications}}
  \textbf{\bibinfo{volume}{10}}, \bibinfo{pages}{2903} (\bibinfo{year}{2019}).
\newblock \urlprefix\url{https://doi.org/10.1038/s41467-019-10827-4}.

\bibitem{Chmiela2018}
\bibinfo{author}{Chmiela, S.}, \bibinfo{author}{Sauceda, H.~E.},
  \bibinfo{author}{Müller, K.-R.} \& \bibinfo{author}{Tkatchenko, A.}
\newblock \bibinfo{title}{Towards exact molecular dynamics simulations with
  machine-learned force fields}.
\newblock \emph{\bibinfo{journal}{Nature Communications}}
  \textbf{\bibinfo{volume}{9}}, \bibinfo{pages}{3887} (\bibinfo{year}{2018}).
\newblock \urlprefix\url{https://doi.org/10.1038/s41467-018-06169-2}.

\bibitem{Chmiela2017}
\bibinfo{author}{Chmiela, S.} \emph{et~al.}
\newblock \bibinfo{title}{Machine learning of accurate energy-conserving
  molecular force fields}.
\newblock \emph{\bibinfo{journal}{Science Advances}}
  \textbf{\bibinfo{volume}{3}}, \bibinfo{pages}{e1603015}
  (\bibinfo{year}{2017}).
\newblock
  \urlprefix\url{http://advances.sciencemag.org/content/3/5/e1603015.abstract}.

\bibitem{Schutt2018}
\bibinfo{author}{Schutt, K.~T.}, \bibinfo{author}{Sauceda, H.~E.},
  \bibinfo{author}{Kindermans, P.~J.}, \bibinfo{author}{Tkatchenko, A.} \&
  \bibinfo{author}{Müller, K.~R.}
\newblock \bibinfo{title}{Schnet - a deep learning architecture for molecules
  and materials}.
\newblock \emph{\bibinfo{journal}{Journal of Chemical Physics}}
  \textbf{\bibinfo{volume}{148}} (\bibinfo{year}{2018}).

\bibitem{Thomas2018}
\bibinfo{author}{Thomas, N.} \emph{et~al.}
\newblock \bibinfo{title}{Tensor field networks: Rotation- and
  translation-equivariant neural networks for 3d point clouds}.
\newblock \emph{\bibinfo{journal}{arXiv preprint arXiv:1802.08219}}
  (\bibinfo{year}{2018}).

\bibitem{Drautz2019}
\bibinfo{author}{Drautz, R.}
\newblock \bibinfo{title}{Atomic cluster expansion for accurate and
  transferable interatomic potentials}.
\newblock \emph{\bibinfo{journal}{Phys. Rev. B}} \textbf{\bibinfo{volume}{99}},
  \bibinfo{pages}{014104} (\bibinfo{year}{2019}).
\newblock \urlprefix\url{https://link.aps.org/doi/10.1103/PhysRevB.99.014104}.

\bibitem{Anderson2019}
\bibinfo{author}{Anderson, B.}, \bibinfo{author}{Hy, T.~S.} \&
  \bibinfo{author}{Kondor, R.}
\newblock \bibinfo{title}{Cormorant: Covariant molecular neural networks}.
\newblock \emph{\bibinfo{journal}{Adv. Neural Inf. Process. Syst.}}
  \textbf{\bibinfo{volume}{32}} (\bibinfo{year}{2019}).

\bibitem{Qiao2020}
\bibinfo{author}{Qiao, Z.}, \bibinfo{author}{Welborn, M.},
  \bibinfo{author}{Anandkumar, A.}, \bibinfo{author}{Manby, F.~R.} \&
  \bibinfo{author}{Miller, I., Thomas~F.}
\newblock \bibinfo{title}{Orbnet: Deep learning for quantum chemistry using
  symmetry-adapted atomic-orbital features}.
\newblock \emph{\bibinfo{journal}{The Journal of Chemical Physics}}
  \textbf{\bibinfo{volume}{153}}, \bibinfo{pages}{124111}
  (\bibinfo{year}{2020}).
\newblock \urlprefix\url{https://doi.org/10.1063/5.0021955}.

\bibitem{Batzner2021}
\bibinfo{author}{Batzner, S.} \emph{et~al.}
\newblock \bibinfo{title}{Se(3)-equivariant graph neural networks for
  data-efficient and accurate interatomic potentials}.
\newblock \emph{\bibinfo{journal}{arXiv preprint arXiv:2101.03164}}
  (\bibinfo{year}{2021}).

\bibitem{Glick2021}
\bibinfo{author}{Glick, Z.~L.}, \bibinfo{author}{Koutsoukas, A.},
  \bibinfo{author}{Cheney, D.~L.} \& \bibinfo{author}{Sherrill, C.~D.}
\newblock \bibinfo{title}{Cartesian message passing neural networks for
  directional properties: Fast and transferable atomic multipoles}.
\newblock \emph{\bibinfo{journal}{The Journal of Chemical Physics}}
  \textbf{\bibinfo{volume}{154}}, \bibinfo{pages}{224103}
  (\bibinfo{year}{2021}).

\bibitem{Schutt2021}
\bibinfo{author}{Sch{\"u}tt, K.}, \bibinfo{author}{Unke, O.} \&
  \bibinfo{author}{Gastegger, M.}
\newblock \bibinfo{title}{Equivariant message passing for the prediction of
  tensorial properties and molecular spectra}.
\newblock In \bibinfo{editor}{Meila, M.} \& \bibinfo{editor}{Zhang, T.} (eds.)
  \emph{\bibinfo{booktitle}{Proceedings of the 38th International Conference on
  Machine Learning}}, vol. \bibinfo{volume}{139} of
  \emph{\bibinfo{series}{Proceedings of Machine Learning Research}},
  \bibinfo{pages}{9377--9388} (\bibinfo{publisher}{PMLR},
  \bibinfo{year}{2021}).
\newblock \urlprefix\url{https://proceedings.mlr.press/v139/schutt21a.html}.

\bibitem{Unke2021}
\bibinfo{author}{Unke, O.~T.} \emph{et~al.}
\newblock \bibinfo{title}{Spookynet: Learning force fields with electronic
  degrees of freedom and nonlocal effects}.
\newblock \emph{\bibinfo{journal}{Nature Communications}}
  \textbf{\bibinfo{volume}{12}}, \bibinfo{pages}{7273} (\bibinfo{year}{2021}).
\newblock \urlprefix\url{https://doi.org/10.1038/s41467-021-27504-0}.

\bibitem{newtonnet_2022}
\bibinfo{author}{Haghighatlari, M.} \emph{et~al.}
\newblock \bibinfo{title}{Newtonnet: a newtonian message passing network for
  deep learning of interatomic potentials and forces}.
\newblock \emph{\bibinfo{journal}{Digital Discovery}}
  \textbf{\bibinfo{volume}{1}}, \bibinfo{pages}{333–343}
  (\bibinfo{year}{2022}).

\bibitem{HCombustion_data_2022}
\bibinfo{author}{Guan, X.} \emph{et~al.}
\newblock \bibinfo{title}{A benchmark dataset for hydrogen combustion}.
\newblock \emph{\bibinfo{journal}{Scientific Data}}
  \textbf{\bibinfo{volume}{9}}, \bibinfo{pages}{215} (\bibinfo{year}{2022}).

\bibitem{wb97x-v_2014}
\bibinfo{author}{Mardirossian, N.} \& \bibinfo{author}{{Head-Gordon}, M.}
\newblock \bibinfo{title}{$\omega${B97X}-{V}: {A} 10-parameter, range-separated
  hybrid, generalized gradient approximation density functional with nonlocal
  correlation, designed by a survival-of-the-fittest strategy}.
\newblock \emph{\bibinfo{journal}{Phys. Chem. Chem. Phys.}}
  \textbf{\bibinfo{volume}{16}}, \bibinfo{pages}{9904--9924}
  (\bibinfo{year}{2014}).

\bibitem{Kulichenko_Barros_2023}
\bibinfo{author}{Kulichenko, M.} \emph{et~al.}
\newblock \bibinfo{title}{Uncertainty-driven dynamics for active learning of
  interatomic potentials}.
\newblock \emph{\bibinfo{journal}{Nature Computational Science}}
  \textbf{\bibinfo{volume}{3}}, \bibinfo{pages}{230–239}
  (\bibinfo{year}{2023}).

\bibitem{shapeev_gubaev_1970}
\bibinfo{author}{Shapeev, A.}, \bibinfo{author}{Gubaev, K.},
  \bibinfo{author}{Tsymbalov, E.} \& \bibinfo{author}{Podryabinkin, E.}
\newblock \bibinfo{title}{Active learning and uncertainty estimation}
  (\bibinfo{year}{1970}).
\newblock
  \urlprefix\url{https://link.springer.com/chapter/10.1007/978-3-030-40245-7_15#Sec8}.

\bibitem{Laio2002}
\bibinfo{author}{Laio, A.} \& \bibinfo{author}{Parrinello, M.}
\newblock \bibinfo{title}{Escaping free-energy minima}.
\newblock \emph{\bibinfo{journal}{Proceedings of the National Academy of
  Sciences}} \textbf{\bibinfo{volume}{99}}, \bibinfo{pages}{12562--12566}
  (\bibinfo{year}{2002}).
\newblock \urlprefix\url{https://doi.org/10.1073/pnas.202427399}.

\bibitem{Barducci2008}
\bibinfo{author}{Barducci, A.}, \bibinfo{author}{Bussi, G.} \&
  \bibinfo{author}{Parrinello, M.}
\newblock \bibinfo{title}{Well-tempered metadynamics: A smoothly converging and
  tunable free-energy method}.
\newblock \emph{\bibinfo{journal}{Phys. Rev. Lett.}}
  \textbf{\bibinfo{volume}{100}}, \bibinfo{pages}{020603}
  (\bibinfo{year}{2008}).
\newblock
  \urlprefix\url{https://link.aps.org/doi/10.1103/PhysRevLett.100.020603}.

\bibitem{Ko2023}
\bibinfo{author}{Ko, T.} \emph{et~al.}
\newblock \bibinfo{title}{Using diffusion maps to analyze reaction dynamics for
  a hydrogen combustion benchmark dataset} (\bibinfo{year}{2023}).
\newblock \eprint{2304.09296}.

\bibitem{Bertels_2020}
\bibinfo{author}{Bertels, L.~W.}, \bibinfo{author}{Newcomb, L.~B.},
  \bibinfo{author}{Alaghemandi, M.}, \bibinfo{author}{Green, J.~R.} \&
  \bibinfo{author}{Head-Gordon, M.}
\newblock \bibinfo{title}{Benchmarking the performance of the reaxff reactive
  force field on hydrogen combustion systems}.
\newblock \emph{\bibinfo{journal}{The Journal of Physical Chemistry A}}
  \textbf{\bibinfo{volume}{124}}, \bibinfo{pages}{5631–5645}
  (\bibinfo{year}{2020}).

\bibitem{Li2004-repeat}
\bibinfo{author}{Li, J.}, \bibinfo{author}{Zhao, Z.}, \bibinfo{author}{Kazakov,
  A.} \& \bibinfo{author}{Dryer, F.~L.}
\newblock \bibinfo{title}{{An updated comprehensive kinetic model of hydrogen
  combustion}}.
\newblock \emph{\bibinfo{journal}{Int. J. Chem. Kinet.}}
  \textbf{\bibinfo{volume}{36}}, \bibinfo{pages}{566--575}
  (\bibinfo{year}{2004}).

\bibitem{Guan_Leven_2021}
\bibinfo{author}{Guan, X.}, \bibinfo{author}{Leven, I.},
  \bibinfo{author}{Heidar-Zadeh, F.} \& \bibinfo{author}{Head-Gordon, T.}
\newblock \bibinfo{title}{Protein c-gem: A coarse-grained electron model for
  fast and accurate protein electrostatics prediction}.
\newblock \emph{\bibinfo{journal}{Journal of Chemical Information and
  Modeling}} \textbf{\bibinfo{volume}{61}}, \bibinfo{pages}{4357–4369}
  (\bibinfo{year}{2021}).

\bibitem{qchem_2014}
\bibinfo{author}{Shao, Y.} \emph{et~al.}
\newblock \bibinfo{title}{Advances in molecular quantum chemistry contained in
  the q-chem 4 program package}.
\newblock \emph{\bibinfo{journal}{Molecular Physics}}
  \textbf{\bibinfo{volume}{113}}, \bibinfo{pages}{184–215}
  (\bibinfo{year}{2014}).

\bibitem{plumed_2014}
\bibinfo{author}{Tribello, G.~A.}, \bibinfo{author}{Bonomi, M.},
  \bibinfo{author}{Branduardi, D.}, \bibinfo{author}{Camilloni, C.} \&
  \bibinfo{author}{Bussi, G.}
\newblock \bibinfo{title}{Plumed 2: New feathers for an old bird}.
\newblock \emph{\bibinfo{journal}{Computer Physics Communications}}
  \textbf{\bibinfo{volume}{185}}, \bibinfo{pages}{604–613}
  (\bibinfo{year}{2014}).

\bibitem{ase_2017}
\bibinfo{author}{Larsen, A.~H.} \emph{et~al.}
\newblock \bibinfo{title}{The atomic simulation environment—a python library
  for working with atoms}.
\newblock \emph{\bibinfo{journal}{Journal of Physics: Condensed Matter}}
  \textbf{\bibinfo{volume}{29}}, \bibinfo{pages}{273002}
  (\bibinfo{year}{2017}).

\bibitem{rupp_2012}
\bibinfo{author}{Rupp, M.}, \bibinfo{author}{Tkatchenko, A.},
  \bibinfo{author}{Müller, K.-R.} \& \bibinfo{author}{von Lilienfeld, O.~A.}
\newblock \bibinfo{title}{Fast and accurate modeling of molecular atomization
  energies with machine learning}.
\newblock \emph{\bibinfo{journal}{Physical Review Letters}}
  \textbf{\bibinfo{volume}{108}} (\bibinfo{year}{2012}).

\bibitem{herman-saffar_2021}
\bibinfo{author}{Herman-Saffar, O.}
\newblock \bibinfo{title}{An approach for choosing number of clusters for
  k-means} (\bibinfo{year}{2021}).
\newblock
  \urlprefix\url{https://towardsdatascience.com/an-approach-for-choosing-number-of-clusters-for-k-means-c28e614ecb2c}.

\bibitem{khaliullin_2007}
\bibinfo{author}{Khaliullin, R.~Z.}, \bibinfo{author}{Cobar, E.~A.},
  \bibinfo{author}{Lochan, R.~C.}, \bibinfo{author}{Bell, A.~T.} \&
  \bibinfo{author}{Head-Gordon, M.}
\newblock \bibinfo{title}{Unravelling the origin of intermolecular interactions
  using absolutely localized molecular orbitals}.
\newblock \emph{\bibinfo{journal}{The Journal of Physical Chemistry A}}
  \textbf{\bibinfo{volume}{111}}, \bibinfo{pages}{8753–8765}
  (\bibinfo{year}{2007}).

\end{thebibliography}
\bibliographystyle{naturemag}


\end{document}


\maketitle

\pagebreak

\begin{table}[htbp]
\begin{center}
\begin{tabular}{ | m{6cm} | m{1cm}| m{1cm} | m{1cm} |}
  \hline 
  No. Reaction & Atoms & $\mathrm{DoF}$ & $\mathrm{DoF_{int}}$ \\ 
  \hline
  \textbf{Association/Dissociation} & & & \\
  5. H$_2\longrightarrow \;$2H & 2 & 6 & 1 \\
  6. O$_2\longrightarrow \;$2O & 2 & 6 & 1 \\
  7. OH$\;\longrightarrow \;$O+H & 2 & 6 & 1\\
  8. H+OH $\longrightarrow$ H$_2$O & 3 & 9 &  3 \\
  9. H+O$_2 \longrightarrow \;$HO$_2$ & 3 & 9 & 3 \\
  15. H$_2$O$_2$ $\longrightarrow\;$2OH & 4 & 12 & 6 \\  
  \hline
  \textbf{Substitution} &&&\\
  16. H$_2$O$_2$+H $\longrightarrow\;$H$_2$O+OH & 5 & 15 & 9 \\
  \hline
  \textbf{O-transfer} &&&\\
  1.  OH+O $\longrightarrow\;$ H+O$_2$ & 3 & 9 & 3 \\
  11. HO$_2$+H$\;\longrightarrow\;$2OH & 4 & 12 &  6 \\
  12. HO$_2$+O$\;\longrightarrow\;$OH+O$_2$ & 4 & 12 &  6 \\ 
  \hline
  \textbf{H-transfer} &&&\\
  2. O+H$_2 \longrightarrow\;$OH+H& 3 & 9 & 3 \\
  3. H$_2$+OH $\longrightarrow\;$H$_2$O+H & 4 & 12 & 6\\
  4. H$_2$O $\longrightarrow\;$2OH & 4 & 12 & 6 \\
  10. HO$_2$+H $\longrightarrow\;$H$_2$+O$_2$ & 4 & 12 & 6 \\
  13. HO$_2$+OH $\longrightarrow\;$H$_2$O+O$_2$& 5 & 12 & 9 \\
  14. 2HO$_2\longrightarrow\;$ H$_2$O$_2$+O$_2$& 6 & 18 & 12 \\
  17. H$_2$O$_2$+H $\longrightarrow\;$HO$_2$+H$_2$& 5 & 15 & 9 \\
  18. H$_2$O$_2$+O $\longrightarrow\;$ HO$_2$+OH & 5 & 15 & 9 \\
  19. H$_2$O$_2$+OH $\longrightarrow\;$H$_2$O+HO$_2$& 6 & 18 & 12 \\
  \hline
\end{tabular}
\end{center}
\caption{The 19 reactions contained in the hydrogen combustion benchmark dataset. The number of atoms involved in each reaction, the total number of degrees of freedom ($\mathrm{DoF}$) in Cartesian coordinates, and total number of degrees of freedom in ICs ( $\mathrm{DoF}_{int}$.)}
\label{tb:reactions}
\end{table}

\begin{table}[htbp]
\begin{center}
\begin{tabular}{|c|c|}
\hline
\textbf{rxn} & \textbf{n data added} \\
\hline
09 & 6686 \\
10 & 6777 \\
13 & 8175 \\
16 & 8234 \\
17 & 8230 \\
18 & 8480 \\
\hline
\end{tabular}
\end{center}
\caption{Total number of data points added in active learning for each reaction.}
\label{tb:n_data_added}
\end{table}

\begin{figure}[htbp]
\center
\includegraphics[width=\textwidth]{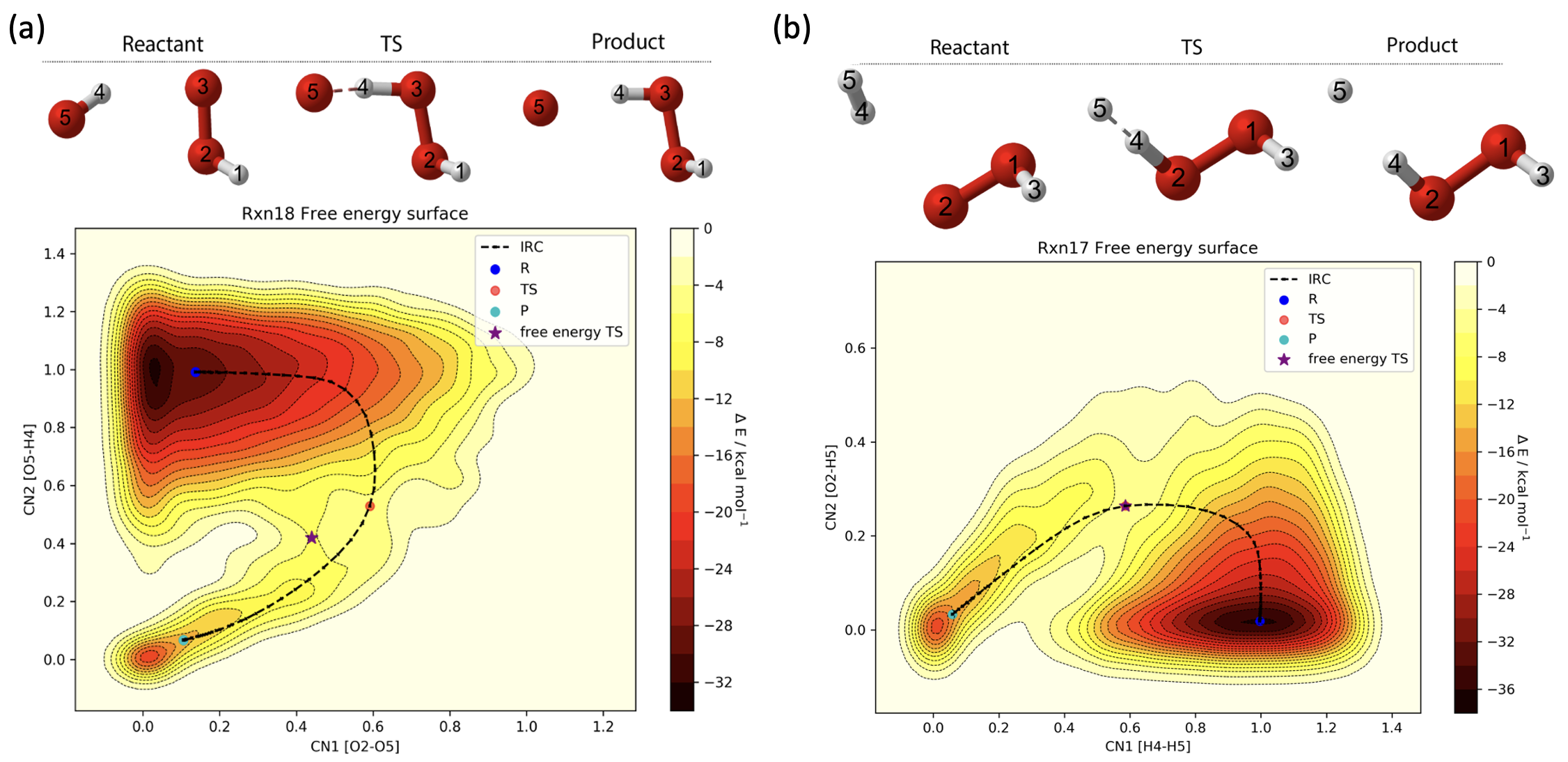}
\caption{Free energy surface reconstructed from metadynamics trajectory of (a) Rxn18 and (b) Rxn17. The original IRC pathway is labeled. The 0K IRC transition state is labeled as the red dot and free energy transition state is labeled as a purple star. For Rxn 18 we can clearly see that the IRC TS on the free energy surface is leaning towards the product side, which nicely explains the 43:57 reactant-to-product ratio in the committer analysis. However a lower transition pathway compared to the original IRC also exists on the free energy surface, with the better free energy transition state residing closer to the reactant. For Rxn17, we see that the IRC transition state is the free energy transition state and hence is enthalpy dominated at 500 K, which also agrees with the 49:51 reactant-to-product ratio from the commitor statistics.  }
\label{fig:FES_SI_18_17}
\end{figure}